\documentstyle[12pt]{article}
\begin{document}
\def\theequation{\arabic{section}.\arabic{equation}}
\begin{titlepage}
\title{MULTIPLE IMAGING \\ BY GRAVITATIONAL WAVES}
\author{Valerio Faraoni \\ \\ {\small \it Department of Physics and Astronomy,
University of Victoria, P.O. Box 3055}\\
{\small \it  Victoria, B.C. Canada V8W 3P6}}
\date{}     \maketitle      \thispagestyle{empty}
\vfill
\begin{abstract} 
Gravitational waves act like lenses for the light propagating 
through them. This phenomenon is 
described using the vector formalism employed for ordinary gravitational 
lenses, which was proved to be applicable also to a non--stationary 
spacetime, with the appropriate modifications. In order to have multiple 
imaging an approximate condition analogous to
that for ordinary gravitational lenses must be satisfied. Certain
astrophysical sources of
gravitational waves satisfy this condition, while the gravitational
wave background, on average, does not. Multiple imaging by gravitational
waves is, in principle, possible, but the probability of observing such 
a phenomenon is extremely low. 
\end{abstract}      \vspace*{1truecm}   \begin{center}
To appear in {\em Int. J. Mod. Phys. D}
\end{center}    \end{titlepage} \clearpage   \setcounter{page}{1}

\section{Introduction} 

In general relativity (and in all metric theories of gravity) a 
gravitational field deflects light rays propagating through it, and
gravitational waves are no exception. It is well known that 
exact solutions of the Einstein equations representing plane gravitational 
waves exhibit focusing properties on timelike and null geodesics propagating 
through them \cite{exactwaves}. Light propagation 
through realistic linearized gravitational waves outside the laboratory 
has also been studied \cite{Zipoy}--\cite{Braginskyetal}. Most authors 
restricted their attention to cosmological gravitational waves; a particular
attention was payed to the frequency shift effect induced by cosmological
gravitational waves on the microwave background photons 
\cite{CMB,Dautcourt,Carr,LiddleLyth}, expecially after the
{\em COBE} discovery of anisotropies in the cosmic microwave background. 
Other works focussed upon the effects of 
gravitational waves perturbing ordinary gravitational lenses 
\cite{McBreenMetcalfe}--\cite{FriemanHarariSurpi}. In recent papers, the 
case of gravitational waves generated by astrophysical sources was studied
\cite{Labeyrie}--\cite{Durrer} and, in some cases, optimistic statements 
were made about the detectability of these waves via the effects 
induced by their interaction with the light coming from distant objects.  

The study of gravitational lensing by 
mass concentrations perturbing the background curvature of the universe has
become a very active field of astronomy \cite{SEF}. In
extreme conditions multiple images of a celestial object are created, which
are accompanied by high amplification events. In these
situations spectacular phenomena are observed by the 
astronomers: multiple quasars, giant arcs, arclets and radio rings 
\cite{SEF}.  It is therefore natural 
to ask whether gravitational waves can create multiple images of a
celestial object and give rise to the associated high amplification events. 
This question was posed in Refs.~\cite{Wheeler}--\cite{Hamburg}. 
The possibility of multiple imaging by gravitational waves in general
relativity constitutes 
the subject of the present paper. Some of the ideas 
exposed here were anticipated in a brief abstract
\cite{Hamburg}. While the idea of multiple imaging by gravitational waves was 
only qualitatively sketched in Ref.~\cite{Wheeler}, in 
Ref.~\cite{BBVarenna} the Raychaudhuri equation was used to deduce 
that the amplification of a light beam
induced by gravitational waves is of second order in the wave amplitudes, and
therefore completely negligible in any situation of astrophysical interest. 
This conclusion is valid only when a single image of the
light source is created, and cannot be extended to the case of multiple images 
and the associated high amplification events because the scalar formalism 
is known to fail in these situations \cite{BlandfordKoch}. Nowadays, 
we have formalisms that are more suitable than the Raychaudhuri equation 
for the description of multiple 
images in gravitational lensing \cite{SEF}. There is, however, a problem 
with the available formalisms, i.e. they were introduced to describe lensing 
in a (conformally) static spacetime and, in principle, their validity is
restricted to this class of spacetimes. Instead, the case considered 
in this paper
involves a highly non--(conformally) stationary spacetime.  This 
problem was resolved in a previous paper \cite{FaraoniApJ},
where it was proved that the vector formalism \cite{SEF} can still be applied 
to the case of lensing by gravitational 
waves, provided that a new formula for the deflection angle is
calculated. This result may look intuitive, 
but the proof of its validity is non--trivial and requires a generalization 
of the Fermat principle to the case of non--stationary spacetimes, which 
was given only recently \cite{Fermat}.
                          
Using the results of Ref.~\cite{FaraoniApJ}, we approach the problem employing 
the vector formalism and describing the effect of the lens with a
plane--to--plane map. The vanishing of the Jacobian of this geometric map 
is associated with the failure of the inverse map, and defines critical 
lines and caustics. An approximate condition for multiple imaging is derived,
and it is shown to be analogous to the well known approximate condition 
for multiple imaging by ordinary gravitational lenses 
\cite{SubramanianCowling}. Order of magnitude estimates
show that the condition for the creation of multiple images by gravitational
waves is satisfied by certain astrophysical sources
of waves considered in the literature. This happens thanks to 
the balance between the (large) source--lens distances and the
(small) wave amplitudes, and because the lensing waves in the sky can have
relatively large amplitudes (i.e. larger than those expected 
in the Solar System). On average, the gravitational wave background is not
capable of multiple imaging. It is concluded that
multiple imaging by gravitational waves is, in principle, possible, but 
the probability of observing such a phenomenon is very low.

Although various aspects of lensing could be considered, in this paper we focus
on the creation of multiple images because the associated high amplification
events are the most dramatic effects, and the most relevant for the detection
of lensing events and, indirectly, of gravitational waves.

The plan of the paper is as follows: in Sec.~2 and Sec.~3 we introduce the
necessary tools and the assumptions underlying this work. Emphasis is given to 
the computation of the deflection angle (Sec.~2) and to the validity of the
thin lens approximation (Sec.~3). In Sec.~4, the modified vector formalism is 
applied to lensing gravitational waves and an approximate condition for the
creation of multiple images is derived. In Sec.~5, gravitational waves and
ordinary gravitational lenses are compared, while Sec.~6 explores the
possibility that the condition for multiple imaging is satisfied by
astrophysical sources of gravitational waves. Sec.~7 contains final remarks.
 
We use units in which the Newton 
constant $ G$ and the speed of light $c$ assume the value unity (but 
$G$ and $c$ will occasionally be restored). The metric
signature is +2; Latin indices run from 1 to 3, and Greek indices run
from 0 to 3. A comma denotes partial differentiation.

\section{Computation of the deflection angle}

Let us consider gravitational waves localized in a region of space between 
a light source and an observer. The spacetime metric is given, in an
asymptotically Cartesian coordinate system $\left\{ t,x,y,z \right\}$, 
by\setcounter{equation}{0}
\begin{equation}
g_{\mu \nu }=\eta_{\mu \nu}+h_{\mu \nu}     \; ,
\end{equation}
where $ \eta_{\mu \nu}=$diag($-1,1,1,1$) and $ |h_{\mu \nu}| \ll 1 $ in these
coordinates. Let us consider a light ray whose unperturbed path is 
parallel to the $ z $--axis. The photon describing this ray has four--momentum
\begin{equation}    \label{usedintheapp}
 p^{\mu}=p^{\mu}_{(0)}+\delta p^{\mu }=(1+\delta p^{0},\delta
p^{1},\delta p^{2},1+\delta p^{3}) \;,
\end{equation}
where $\delta p^{\mu } $ are small deflections (of order $ h_{\mu \nu }
$) and the unperturbed photon four--momentum is $ p_{(0)}^{\mu }=(1,0,0,1) $. 
The computations are performed in the geometric optics approximation, 
that holds if the wavelength $\lambda_{gw} $ of the gravitational wave is 
much larger than the photon wavelength $ \lambda_{em} $, and if
\begin{equation}     \label{3}
\lambda_{em} >{\lambda_{gw}}^2 /D_{L}  \; ,
\end{equation}
where $ D_{L}$ is the observer--lens distance. Equation~(\ref{3}) 
guarantees that the size
of the interference fringes which eventually form at the observer's position 
is not comparable to the ``geometrical shadow''
of the lens \cite{BBVarenna}. In order to make computations to
second order in the wave amplitudes, tensor indices are lowered and raised 
with $ g_{\mu \nu} $ and with
\begin{equation}
g^{\mu \nu }=\eta^{\mu \nu }-h^{\mu \nu }+O(h^{2}) \equiv \eta^{\mu \nu }-
\eta^{\mu \rho }\eta^{\nu \sigma }h_{\rho \sigma }+O(h^{2}) \;,
\end{equation}
respectively. The symbols $O(h)$, $O(h^2)$ denote the orders of magnitude of
$h_{\mu\nu}$ and of terms quadratic in $h_{\mu\nu}$, respectively. 
The equation of null geodesics gives
\begin{equation}
\frac{d( \delta p^{\mu })}{d\lambda }+\Gamma^{\mu}_{\rho \sigma}\left(
p_{(0)}^{\rho}+\delta p^{\rho} \right) \left( p_{(0)}^{\sigma}+\delta
p^{\sigma} \right)=0 \;,
\end{equation}
where $ \lambda $ is an affine parameter along the null geodesics and
\begin{equation}
\Gamma ^{\mu}_{\rho \sigma}=\frac{1}{2}\,g^{\mu \nu} \left(
h_{\nu \rho ,\sigma}+h_{\nu \sigma ,\rho }-h_{\rho \sigma ,\nu } \right)
 \; .                \end{equation}
One obtains 
\begin{equation} \label{geodesicequation}
\frac{ d( \delta p^{\mu})}{d\lambda}+\eta^{\mu \nu} \left(
h_{\nu 0,0}+h_{\nu 0,3}+h_{\nu 3,0}
+h_{\nu 3,3}
-h_{03,\nu}-\frac{1}{2}\,h_{00,\nu}-
\frac{1}{2}\,h_{33,\nu} \right) +{\mbox O}(h^2) =0 \; .
\end{equation}
We introduce the notation
\begin{equation}
A_{\lambda} \equiv p^{\alpha} A_{\alpha} \; ; \;\;\;\;
B_{,\lambda} \equiv p^{\alpha}\frac{\partial B}{\partial x^{\alpha}} 
\;.    \end{equation}
Using the expression
\begin{equation}
 \Gamma^{\mu }_{\lambda \lambda}=\eta^{\mu \nu} \left(
h_{\nu \lambda,\lambda}-\frac{1}{2}\,\,h_{\lambda \lambda ,\nu} \right)
+O(h^{2})    \end{equation}
and Eq.~(\ref{geodesicequation}), one obtains
\begin{equation}
\delta p^{\mu}=-\int_{S}^{O} d\lambda\,\,\left(
{h^{\mu}}_{\lambda,\lambda}-\frac{1}{2}\,{h_{\lambda \lambda}}^{,\mu} \right)
+O(h^{2}) \;,
\end{equation}
where the indices are now raised with $ \eta^{\mu \nu} $ and the integral
is computed along the photon path from the source to the observer. Since 
a {\em localized} pulse of gravitational waves is considered, the first 
term in the integrand gives a vanishing boundary term\footnote{The 
boundary term is usually negligible \cite{Durrer}. The case of ``pulses 
with memory'' was considered in Ref.~\cite{FakirApJ93}.} 
and it follows that 
\begin{equation} \label{deflections}
\delta p^{\mu}=\,
\frac{1}{2}\,\int_{S}^{O}d\lambda \,\,\left(
h_{00}+2h_{03}+h_{33} \right)^{,\mu} +O(h^{2}) \;.
\end{equation}
Performing the integration along the {\em unperturbed} photon path
instead of the actual path involves only an error of order $ h^{2} $ and 
hence the first order result is given by
\begin{equation}  \label{deflectionangle}
\delta p_{(1)}^{\mu}=\, \frac{1}{2} \,\int_{S}^{O} dz\,\,\left(
{h_{00}}+2{h_{03}}+{h_{33}} \right)^{,\mu} 
\end{equation}
(notice that the deflection angle $\delta p^{\mu}$ is the opposite of 
the angle used in the theory of ordinary gravitational lenses, in which
deflection from a straight line is
defined as having the opposite orientation).
For pulses of gravitational waves this integral is extended to a region of
space of order $ P $ (the characteristic period of the waves) where the
pulse is localized so that, in order of magnitude, $ \delta p \sim
P\cdot h/P =h $. If the gravitational wave background is considered, it
might appear that the computation of the integral along the whole photon 
path from the source to the observer gives $ \delta p \sim D h/P $. However, 
this secular effect is absent, i.e. the deflection does not cumulate with 
the travelled
distance, due to the transversality of gravitational waves and to the
equality between the speed of light and that of the random
inhomogeneities of the medium (the gravitational waves), as shown in 
Refs.~\cite{Zipoy,ZipoyBertotti,Dautcourt,BertCat,Linder,Braginskyetal}. 

We also compute the second order correction $ \delta p^{\mu}_{(2)} $ to the 
deflection, which was not given in previous references and will be used in
Sec.~4. $ \delta p^{\mu}_{(2)} $ is obtained by inserting 
Eq.~(\ref{deflectionangle}) into the equation of null geodesics
\begin{equation}
\frac{d( \delta p^{\mu} )}{d\lambda}+\Gamma_{\lambda \lambda}^{\mu}+
2\Gamma_{\lambda \sigma}^{\mu}\delta p_{(1)}^{\sigma}+O(h^{3})=0 \;,
\end{equation}
where $ \delta p^{\mu}=\delta p_{(1)}^{\mu}+\delta p_{(2)}^{\mu} $,
which gives
\begin{eqnarray}
\delta p_{(2)}^{\mu}=\int_{S}^{O}dz \,\, \left[ -\left(
\partial_{\sigma}h^{\mu}_{0}+\partial_{\sigma}h^{\mu}_{3}+
\partial_{0}h^{\mu}_{\sigma}+\partial_{3}h^{\mu}_{\sigma}-
\partial^{\mu}h_{0\sigma}-\partial^{\mu}h_{3\sigma} \right) \delta
p_{(1)}^{\sigma}  \right. \nonumber  \\
\mbox{} \left. 
+h^{\mu\alpha} \left( h_{\alpha 0,0}+h_{\alpha 0,3}+h_{\alpha 3,0}
+h_{\alpha 3,3}-h_{03,\alpha}-\frac{1}{2}h_{00,\alpha}
-\frac{1}{2} h_{33,\alpha} \right) \right] +O(h^{3})  \;,
\label{deflectionangle2}
\end{eqnarray}
with $ \delta p_{(1)}^{\sigma} $ given by Eq.~(\ref{deflectionangle}) and
where the integration is performed again along the unperturbed
photon's path instead of the actual path, the difference contributing
only by a third order term. Equation~(\ref{deflectionangle2}) allows the 
computation of the second order contribution to the divergence 
$\partial_{A}( \delta p^{A} )$ (the first order contribution being zero
\cite{BBVarenna}):
\begin{eqnarray}
\partial_{A}( \delta p^{A} )=\int_{S}^{O}dz\, \left\{ -\delta p_{(1)}^{\sigma}
\left( {h^{A}}_{0,\sigma A}+{h^{A}}_{3,\sigma A}+{h^{A}}_{\sigma ,0A}+
{h^{A}}_{\sigma ,3A} \right)  \right. \nonumber  \\
\mbox{} \left. -\partial_{A} \left( \delta p_{(1)}^{\sigma} \right) \left(
{h^{A}}_{0,\sigma}+
{h^{A}}_{3,\sigma}+{h^{A}}_{\sigma ,0}+{h^{A}}_{\sigma ,3}-{h_{0\sigma}}^{,A}-
{h_{3\sigma}}^{,A}\right)  \right. \nonumber \\
\mbox{} \left. +{h^{A \alpha}}_{,A} \left( h_{\alpha 0,0}+ h_{\alpha 0,3}
+ h_{\alpha 3,0}+ h_{\alpha 3,3}- h_{03,\alpha}-\frac{1}{2} h_{00,\alpha}
-\frac{1}{2} h_{33,\alpha} \right) \right. \nonumber \\
\mbox{} \left. +h^{A\alpha} \left(  h_{\alpha 0,0A}+ h_{\alpha 0,3A}+ 
h_{\alpha 3,0A}+ h_{\alpha 3,3A}- h_{03, \alpha A}-\frac{1}{2} 
h_{00,\alpha A}- \frac{1}{2} h_{33,\alpha A} \right) \right\}     
\; , \label{J1}
\end{eqnarray}
where $A=1,2$.

\section{The thin lens approximation}

In the next section we will apply the vector formalism for gravitational 
lenses,
with the appropriate modifications, to the case of lensing by gravitational
waves. One of the assumptions underlying the vector formalism is the 
validity of the thin lens approximation, i.e. that the size of the lens be 
negligible in
comparison to the length of the path travelled by the photon. This
approximation requires special care in the case of lensing
gravitational waves, and is dealt with in what follows.

To be specific, we restrict to the case of gravitational waves generated by
astrophysical sources and propagating with wavefronts that are approximately
spherical at a sufficiently large distance from the source. In this 
situation, a
photon coming from infinity and entering the gravitational wave will never
leave it, hence the thin lens approximation might seem to be 
inapplicable. However, the photon is appreciably deflected only during 
a small fraction of the time spent
inside the gravitational wave, at the minimum distance from the 
source of gravitational waves (where the wave amplitude is maximum). 
The region of space in which the photon is appreciably 
deflected has a size that is
much smaller than the total length of the path travelled by the photon. In
fact, let $h \sim h_S R_S/r$ be the order of magnitude of the gravitational
wave amplitude at a distance $r$ from the source of waves, where $R_S$ and
$h_S$ are, respectively, the amplitude at the source and the Schwarzschild
radius of the object generating gravitational waves. One has $h_S \simeq
\epsilon \, M/R$, where $\epsilon$, $M$ and $R$ are the efficiency of the
generation process, the mass and radius of the object, respectively.
For neutron stars, $M/R\sim 1/20$, and $\epsilon \sim 0.1$ is a very generous
estimate of the efficiency (see Sec.~6). These numbers give a 
deflection angle $\delta \sim
h \sim 5\cdot 10^{-3}$~(1~km$/r$). The minimum detectable separation between
multiple images is of the order $\delta_0 \simeq 10^{-4}$ arcseconds,
marginally accessible by VLBI. The requirement $\delta \geq \delta_0$ gives
$r\sim h_S R_S/\delta \leq 3\cdot 10^7$~km. This distance is very small in
comparison with the path travelled by the photon, usually of the order of many
megaparsecs. Allowing a separation between the images of a light source
smaller than $\delta_0 \sim 10^{-4}$~arcseconds would still
give high amplification events although it would be impossible to resolve 
the multiple images (this occurs, e.g. in microlensing by stars 
or compact objects \cite{SEF}). Even in this case the size of the region 
where the deflection takes place is negligible in comparison to the size 
of the photon path.

A second argument supporting the validity of the thin lens 
approximation is the
following: Due to the transversality of gravitational waves, a photon is {\em 
not} deflected by a gravitational wave propagating parallely to the photon 
path. To
be more specific, let $\vec{k}$ be the (3-dimensional) wave vector of a 
gravitational wave of finite (short) duration. If
the component $k^3$ is zero (i.e. the gravitational wave propagates
orthogonally to the unperturbed photon path), it is clear that the
gravitational wave is confined to a plane and that the thin lens 
approximation holds. 
One may conjecture that this approximation fails to be satisfied as $|k^3|$
increases, but this would be incorrect. We give an explicit proof for 
a single monochromatic gravitational 
wave. In the extreme case $ |\vec{k}|=|k^3|$
(gravitational wave antiparallel\footnote{If the gravitational wave propagates
parallely to the light rays, the light never reaches the gravitational wave,
due to the equality between the speed of gravitational waves $c_{gw}$ and the 
speed of light. This situation may change in alternative theories of gravity 
in which $c_{gw}\neq c$ \cite{Will}.} to the photon path) the 
light rays are not
deflected, due to the transversality of gravitational waves. This
fact can be deduced by Refs.~\cite{Zipoy,Linder} or by 
Eq.~(\ref{deflectionangle}). In fact, the use of the
transverse--traceless (TT) gauge in Eq.~(\ref{deflectionangle}) gives
$h_{00}^{(TT)}=h_{03}^{(TT)}=h_{33}^{(TT)}=0$, which implies
$\delta p^A=0$. The vanishing of the deflection angle (a gauge--invariant 
quantity) in the TT gauge implies its vanishing in any gauge.

If the direction of propagation of the gravitational wave 
and the (unperturbed)
photon path are not parallel nor orthogonal, the light rays 
can be approximated
by zig--zag paths with (3--dimensional) unit tangent vector 
$\vec{n}$, and the deflection
depends on the quantity $( h_{00}+2h_{0i}n^i+h_{ij}n^i n^j )$, which replaces
the argument of the integral in Eq.~(\ref{deflectionangle}) \cite{FaraoniApJ}. 
The vector $\vec{n}$ is decomposed into its components parallel and
perpendicular to $\vec{k}$:\setcounter{equation}{0}
\begin{equation}
\vec{n}=\vec{n}_{\parallel}+\vec{n}_{\perp} \; ,
\end{equation}
where $\vec{n}_{\parallel}=\vec{k} \left( \vec{n}\cdot \vec{k}/k^2 \right) $ 
and $\vec{n}_{\perp}\cdot \vec{k}=0$. By introducing the quantities
\begin{equation}
h_{0 \,\parallel} \equiv h_{0i} \, n^i_{\parallel} \;\;\; , \;\;\; 
h_{0\perp} \equiv h_{0i}\, n^i_{\perp} \; ,
\end{equation}
\begin{equation}
h_{\parallel \; \parallel} \equiv h_{ij}\, n^i_{\parallel}\, 
n^j_{\parallel} \;\;\; ,\;\;\;
h_{\parallel \; \perp} \equiv h_{ij}\, n^i_{\parallel}\, 
n^j_{\perp} \;\;\; , \;\;\;
h_{\perp \; \perp} \equiv h_{ij}\, n^i_{\perp} n^j_{\perp} \; ,
\end{equation}
one can write
\begin{equation}
h_{00}+2h_{0i}\,n^i+h_{ij}\,n^i n^j=
h_{00}+2\left( h_{0\parallel}+h_{0\perp}+h_{\perp \; \parallel} \right) 
+h_{\perp \; \perp}
+h_{\parallel \; \parallel} \; .
\end{equation}
The monochromatic gravitational wave is best described in the
TT gauge in which $h_{0\mu}=h_{\parallel \;
\parallel}=h_{\parallel \; \perp}=0$. In this gauge
\begin{equation}
h_{00}+2\left( h_{0\, \parallel}+h_{0\perp}+h_{\perp \; \parallel}\right)
+h_{\perp \; \perp}+h_{\parallel \; \parallel}=h_{\perp \; \perp} \; .
\end{equation}
Therefore, the ``component'' of the gravitational wave propagating parallely 
to the (unperturbed) photon path does not affect the light rays, and the 
thin lens approximation is valid.

\section{The vector formalism and the occurrence of multiple images} 

The vector formalism familiar from the description of ordinary
gravitational lenses can be employed also for the case in which the lens is a 
gravitational wave, provided that the deflection angle (\ref{deflectionangle}) 
is used. The proof of this statement was presented in a previous paper
\cite{FaraoniApJ}, and relies on a rigorous formulation 
of the Fermat principle
valid for arbitrary (non--stationary) spacetimes which became available only
recently \cite{Fermat}.

We set the geometry of the problem as customary in gravitational lens
theory: we consider only waves which are localized near a plane $ z=z_{w}
$ to satisfy the thin lens approximation; the deflection takes 
place in this plane ({\em lens} or {\em image plane} in the usual
language) \cite{SEF}. This description is clearly adequate for the case of 
a gravitational wave propagating in a direction orthogonal to the $z$--axis.  
Naively, one might expect that the description becomes less and 
less correct as the
direction of propagation of the gravitational wave becomes closer and 
closer to the $z$--axis, but this is not the case, as discussed in 
the previous section.

Let $ \underline{x}=(x,y) $ be the
apparent source position in the lens plane, and 
$ \underline{s}=(s_{x},s_{y}) $
be the true source position
(i.e. its position were the lensing wave absent). The plane orthogonal to 
the $ z $--axis and
passing through the source is referred to as the {\em source plane}. The 
true source position in the lens plane is related to that in the source 
plane by identifying the two planes). The
deflection is described by a two--dimensional vector field
$ \delta p^{A}( \underline{x} ) $ ($ A=1,2 $) in the lens plane. 
The action of the
lens is described by a plane--to--plane mapping $ x^{A} \longmapsto
s^{A} $, where $ \underline{s} $ is given by the {\em lens 
equation}\setcounter{equation}{0}
\begin{equation}  \label{lensequation}
s^{A}=x^{A}+\frac{D_{L}D_{LS}}{D_{S}}\,\, \delta p^{A}( \underline{x} ) 
\end{equation}
and where $ D_{L} $, $ D_{LS} $ and $ D_{S} $ are the observer--lens,
lens--source, and observer--source distances, respectively. As customary in
gravitational lens theory, we can fit cosmology into the model by
assuming the $D$'s to be angular diameter distances in a
Friedmann--Lemaitre--Robertson--Walker (FLRW) universe \cite{SEF}.
However, for the sake of simplicity,  we assume that the background is 
flat and that the $ D$'s 
denote Euclidean distances, with $D_{S}=D_{L}+D_{LS} $.
As described in Ref.~\cite{FaraoniApJ}, the use of the Fermat principle 
generalized to non--stationary spacetimes \cite{Fermat} allows one to derive 
the lens equation ~(\ref{lensequation}), with 
the deflection angle $\delta p^A$
given by Eq.~(\ref{deflectionangle}).

The map described by Eq.~(\ref{lensequation}) has the Jacobian matrix
\begin{equation}
J\left( \begin{array}{c}
\underline{s} \\ \underline{x}
\end{array}   \right) =\left( \frac{\partial s^{A}}{\partial x^{B}} \right)
=\left( \begin{array}{cc}
1+D \,\partial_{x}( \delta p^{x})  & D\, \partial_{y}( \delta p^{x}) \\
D\,\partial_{x}( \delta p^{y})   & 1+D\,\partial_{y} ( \delta p^{y})
\end{array}  \right) \; ,
\end{equation}
where $ D\equiv D_{L}D_{LS}/D_{S} $. The inverse matrix $ A=J^{-1} $
represents the {\em amplification tensor}, while its determinant $ {\cal
A}=\mbox{Det}(J)^{-1} $ is the (scalar) {\em amplification}. Since the 
surface
brightness is conserved during lensing whenever geometric optics holds
\cite{Etherington} and
\begin{equation}
 {\cal A}=\frac{area\,\,\,\,of
\,\,\,\,an\,\,\,\,infinitesimal\,\,\,\,region\,\,\,\,in\,\,\,\,the\,\,\,\,
lens\,\,\,\,plane}{area\,\,\,\,of\,\,\,\,the\,\,\,\,corresponding\,\,\,\,
region\,\,\,\,in\,\,\,\,the\,\,\,\,source\,\,\,\,plane} \:\:,
\end{equation}
$\cal A $ has also the meaning of the ratio of light intensities with and
without the lens \cite{SEF}. A small circular source will be imaged into
a small ellipse whose eccentricity $ e $ is given by the ratio of the
eigenvalues $ e_{\pm} $ of $ A $:
\begin{equation}
 (1-e^{2})^{1/2}=\left| \frac{e_{+}}{e_{-}}\right| \;.
\end{equation}
The vanishing of the Jacobian $ \mbox{Det}(J) $ indicates the failure of
invertibility of the map~(\ref{lensequation}). The loci of points in the
lens plane where $ \mbox{Det}(J)=0 $ are called {\em critical lines} and the
corresponding curves in the source plane are called {\em caustics}. Critical 
lines
[caustics] separate regions corresponding to different numbers of images in
the lens [source] plane. Therefore, {\em the occurrence of multiple images is
signalled by the vanishing of} Det$( J)$, and this is the condition that we
will study in the following. 
These concepts are familiar from standard gravitational lens theory 
\cite{SEF}. We now proceed to study the features that are peculiar to lensing
gravitational waves.
 
Let us consider the Jacobian determinant
\begin{equation}   \label{determinant}
\mbox{Det}(J)=1+D\,\,\frac{\partial ( \delta p^{A})}{\partial x^{A}}-
D^{2}\left[ \partial_{y}( \delta p^{x}) \cdot \partial_{x}( \delta p^{y})-
\partial_{x}( \delta p^{x}) \cdot \partial_{y}( \delta p^{y}) \right]
\; ;
\end{equation}
the divergence $ \partial ( \delta p^{A}) /\partial x^{A} $ vanishes
to first order. This follows from the fact that it represents the expansion
scalar of a congruence of null rays. It was proved correctly in 
Ref.~\cite{BBVarenna} 
that this quantity is of second order in the wave amplitudes; we give an
independent proof in the Appendix.  
Eq.~(\ref{determinant}) can be written as follows
\begin{equation}
\mbox{Det}(J)=1+J_1+J_2 \; , \end{equation}
\begin{equation}
J_1=\sqrt{f( \alpha )}\,D_{S}\,
\frac{\partial ( \delta p^{A})}{\partial x^{A}} \;,    \end{equation}
\begin{equation}
J_2=f( \alpha )D_{S}^{2} \left[
\partial_{x}( \delta p^{x}) \cdot \partial_{y}( \delta p^{y})-
\partial_{y}( \delta p^{x}) \cdot \partial_{x}( \delta p^{y}) \right] \; ,
\end{equation}
where $ \alpha \equiv D_{LS}/D_{S}  $ and $ f( \alpha
)=\alpha^{2}(1-\alpha )^{2} $. The polynomial $f( \alpha ) $ is symmetric
about $ \alpha =1/2 $ (corresponding to $ D_{L}=D_{LS} $), where it
assumes its maximum value $ 1/16\simeq 0.0625 $. 

From Eq.~(\ref{J1}) one obtains the order of magnitude estimate
\begin{equation}   \label{estimate}
J_{1}\sim D\, \partial_{A}( \delta p^{A}) \sim \frac{D}{P} \,h^{2} \; .
\end{equation}
Since $ J_{2} \sim (D/P)^{2} h^{2} $, we neglect $ J_{1} $ in
comparison to $ J_{2} $ in the Jacobian determinant of 
Eq.~(\ref{determinant}) when large values of $ D/P $ are considered, which is
the case of interest.

In order for
$ \mbox{Det}(J) $ to vanish (i.e. to have multiple
images of the light source), it must be $ J_{1}+J_{2}<0 $. Since $ J_{1} $ 
and $ J_{2} $ are
terms of order $ h^{2} $ times $ D/P $ or $ (D/P)^{2} $, 
respectively (see Eqs.~(\ref{deflectionangle}) and (\ref{estimate})), 
in order to have $ \mbox{Det}(J)=0 $, a large value of the distance
$ D $ must balance for the small values of the wave amplitudes. Moreover, 
for large
values of $ D/P $, $ J_{1} $ is negligible in comparison with $ J_{2}$, 
and hence in order to have multiple imaging it must be 
\begin{equation}   \label{lab}
f( \alpha )\left[ D_{S}\, \partial_{A}( \delta p^{B}) \right] ^{2}\sim 1
\;.
\end{equation}
For standard gravitational lenses, the probability of lensing of a distant
source is approximately maximum when the lens is halfway between the source 
and the observer \cite{TOG}. This applies also to lensing gravitational waves 
and therefore values of the  numerical factor
$ f( \alpha ) $ far from its maximum are not statistically
significant. We take $ f( \alpha ) $ in the range $ \frac{1}{100} 
$~--~$\frac{1}{16} $ near its maximum. Then, in order to have multiple
imaging, the following condition must be satisfied:
\begin{equation}    \label{inequality}
D\, \frac{\delta p}{P}\sim  \frac{Dh}{P} \geq 4-10 \;,
\end{equation}
where $ P $ is the period of the gravitational wave (computed at
the redshift of the lens plane, if the $ D's $ are chosen to represent
angular diameter distances in a FLRW universe) and
where we used $ \delta p\sim h $ (see Eq.~(\ref{deflectionangle})). 
The inequality (\ref{inequality}) can be written in the form
\begin{equation}  \label{roughcondition}
\frac{h}{P}\geq {\cal S}_{c} \;,
\end{equation}
where ${\cal S}_{c}\equiv (4-10)c/D $. The approximate condition for multiple
imaging~(\ref{roughcondition})
involves the ``strength'' $ h $ and the ``size'' $ P $ of gravitational waves,
the geometry of the problem (through $ D $) and the fundamental constant
$ c $. (\ref{roughcondition}) is analogous to the well known
condition for multiple imaging by ordinary gravitational lenses
\cite{SubramanianCowling}
\begin{equation}   \label{roughgl}
\Sigma \geq \Sigma_{c}\equiv \frac{c^{2}}{4\pi GD}  \;,
\end{equation}
where $ \Sigma $ is the two--dimensional (projected or surface) density 
of the lens and
the {\em critical density} $ \Sigma_{c} $ depends only from the
geometrical factor $ D $ and from fundamental constants. This is a
condition on the ``strength'' (the mass) and the size of the lens. 

The energy and momentum of high frequency gravitational waves are given 
by the Isaacson effective stress tensor which, in the TT gauge, assumes  
the form         
\begin{equation} \label{Isaacson}
T_{\mu\nu}=\frac{c^4}{32\pi G}\, \langle \sum_{i,j} h_{ij,\mu}^{(TT)} \, 
h_{ij,\nu}^{(TT)}  \rangle \; ,
\end{equation}
where $\langle$~$\rangle$ denotes the Brill--Hartle average. 
For $\mu=\nu=0$, $T_{00}=\rho c^2$, where $\rho$ is the effective mass
density of the gravitational waves. It is clear from this expression that 
the ratio $h/P$ is a measure of the ``mass 
density'' of the particular lens under consideration\footnote{Although 
the form
of the perturbations $h_{\mu\nu}$ depends on the chosen gauge, its order of
magnitude, and that of $h/P$, is unaffected by gauge transformations
\cite{FaraoniPLA}. Therefore the approximate condition (\ref{roughcondition}) 
for multiple imaging is gauge--independent.}.
We comment on the analogy between eq.~(\ref{roughcondition}) and 
eq.~(\ref{roughgl}): By squaring, dividing by $G$ and integrating 
eq.~(\ref{roughcondition}), it is deduced that
\begin{equation}     \label{ref2}
\Sigma_{gw} \equiv \frac{1}{G} \int_{S}^{O}dz \left( \frac{h}{P} 
\right)^2 \geq {S_c}^2 \frac{D_S}{G}
\simeq ( 16-100) \frac{c^2}{GD} \; .
\end{equation}
Strictly speaking, it is eq.~(\ref{ref2}), not (\ref{roughcondition}) that 
is the analogous of eq.~(\ref{roughgl}): the quantity $\Sigma_{gw}$ 
has the dimensions of a surface density, and $S_c ^2 D_S/G$ is the critical 
surface density.

The particular lens under consideration always produces an odd number of
images. In fact, the odd image number theorem \cite{Burke} holds, its proof
requiring only boundedness, smoothness and transparency of the lens 
(conditions satisfied by localized gravitational waves). 

\section{Comparison between a gravitational wave and an ordinary gravitational
lens}
 
It is instructive to compare the action of a gravitational wave with that
of an ordinary gravitational lens. The latter is a mass distribution described
by a Newtonian potential $ \Phi $ (satisfying the Poisson equation $
\nabla^{2} \Phi =4\pi \rho $, where $ \rho $ is the lens mass density). It is 
assumed that the lens is smooth, bounded and stationary, i.e
\begin{itemize}
\item $ \Phi $ is continuous with its first and second derivatives;
\item $ \Phi \rightarrow 0 $ and $ \nabla \Phi \rightarrow 0 $ as $
r\equiv \left( x^{2}+y^{2}+z^{2} \right) ^{1/2} \rightarrow +\infty $;
\item $ \partial \Phi /\partial t \simeq 0 $.
\end{itemize}
The plane--to--plane map describing the lens action is given by the lens
equation and the Jacobian matrix can be written \cite{BlandfordKoch}
as\setcounter{equation}{0}
\begin{equation}
J=\left( \begin{array}{cc}
1-\chi -\Lambda    &      -\mu \\
-\mu               &      1-\chi +\Lambda
\end{array}   \right) \;,
\end{equation}
with
\begin{eqnarray}
\chi & \equiv & \frac{\Sigma}{\Sigma_{c}} \;, \\
\Lambda & \equiv & \frac{D}{c^{2}} \int_{-\infty}^{+\infty} dl\, \left(
\frac{\partial^{2} \Phi }{\partial x^{2}}-
\frac{\partial^{2} \Phi }{\partial y^{2}}   \right)  \;,  \\
\mu & \equiv & \frac{D}{c^{2}} \int_{-\infty }^{+\infty} dl \,\,
\frac{\partial^{2}\Phi }{\partial x\partial y} \; ,
\end{eqnarray}
where $ D $ and $\Sigma_c$ have been defined in the previous section and 
\begin{equation}
\Sigma \equiv \int_{-\infty}^{+\infty}dl\, \rho  \; .
\end{equation}
The Jacobian determinant is given by
\begin{equation}
\mbox{Det}(J)=(1-\chi )^{2}-( \Lambda^{2}+\mu^{2})  \;.
\end{equation}
The {\em convergence} $ \chi $ describes the action of matter, while $
\Lambda $ and $ \mu $ describe the action of shear.
For a lensing gravitational wave one obtains 
\begin{equation}
J_{gw}=\left(   \begin{array}{cc}
1-\Lambda_{1}  &    -\mu_{1}   \\
-\mu_{2}       &    1-\Lambda_{2}
\end{array}   \right) \;,
\end{equation}
where
\begin{equation}
\Lambda_1=-D \partial_x \left( \delta p^x \right)=
-D \partial_x \left( \delta p^x_{(1)}+\delta p^x_{(2)}\right) 
\equiv \Lambda_{gw}+\delta \Lambda_1 \; ,
\end{equation}
\begin{equation}
\Lambda_2=-D \partial_y \left( \delta p^y \right)=
-D \partial_y \left( \delta p^y_{(1)}+\delta p^y_{(2)}\right) 
\equiv -\Lambda_{gw}+\delta \Lambda_2 \; ,
\end{equation}
\begin{equation}
\mu_1=-D \partial_y \left( \delta p^x \right)=
-D \partial_y \left( \delta p^x_{(1)}+\delta p^x_{(2)}\right) 
\equiv \mu_{gw}+\delta \mu_1 \; ,
\end{equation}
\begin{equation}
\mu_2=-D \partial_x \left( \delta p^y \right)=
-D \partial_x \left( \delta p^y_{(1)}+\delta p^y_{(2)}\right) 
\equiv \mu_{gw}+\delta \mu_2  \; .
\end{equation}
To first order, one has
\begin{eqnarray}  \label{landa1}
\Lambda_{1} & \equiv & -\frac{D}{2} \int_{S}^{O} d\lambda \,\,
{{h_{\lambda \lambda }}^{,x}}_{,x} \;, \\
\label{landa2}
\Lambda_{2} & \equiv & -\frac{D}{2}\int_{S}^{O} d\lambda \,\,{{h_{\lambda
\lambda}}^{,y}}_{,y} \;,  \\
\label{mu1}
\mu_{1} & \equiv & -\frac{D}{2} \int_{S}^{O} d\lambda \,\,
{{h_{\lambda \lambda }}^{,x}}_{,y} \;,  \\
\label{mu2}
\mu_{2} & \equiv & -\frac{D}{2}\int_{S}^{O} d\lambda \,\,
{{h_{\lambda \lambda}}^{,y}}_{,x} \;,
\end{eqnarray}
where the integrals are computed along the photon's path from the source
to the observer. To first order, 
$ \Lambda_{2}=-\Lambda_{1}\equiv -\Lambda_{gw}
$ and $ \mu_{1}=\mu_{2}\equiv \mu_{gw} $. In fact, from 
Eqs.~(\ref{landa1})--(\ref{mu2}) and from $\partial_A \left( 
\delta p^A \right)=0$ (see the Appendix), it follows that 
\begin{eqnarray}
\Lambda_{1}+\Lambda_{2}=-\frac{D}{2}\,\int_{S}^{O} d\lambda \,\,\left(
{{h_{\lambda \lambda }}^{,x}}_{,x}+
{{h_{\lambda \lambda }}^{,y}}_{,y} \right) +O(h^{2})= & & \nonumber \\
\mbox{} = -\frac{D}{2}\,\int_{S}^{O} d\lambda \,\,p^{\alpha}p^{\beta}\left(
{{h_{\alpha \beta }}^{,x}}_{,x}+
{{h_{\alpha \beta }}^{,y}}_{,y} \right) +O(h^{2}) & & =0+O(h^{2}) \; , 
\end{eqnarray}
hence $ \Lambda_{2}=-\Lambda_{1}+O(h^{2}) $. 

In the first order approximation, 
tensor indices are raised and lowered with $ \eta^{\mu
\nu} $ and $ \eta_{\mu \nu } $, respectively, and $ 
\eta^{AB}=\eta_{AB}=\delta_{AB} $ for
$A,B=1,2 $. Hence, to this order, $ {{h_{\mu \nu }}^{,A}}_{,B}=
{{h_{\mu \nu }}^{,B}}_{,A} $ and $ \mu_{1}=\mu_{2}+O(h^{2}) $.

The corrections 
\begin{equation} 
\delta \Lambda_1=-D\partial_x \left( \delta p_{(2)}^x \right) \; ,
\end{equation}
\begin{equation} 
\delta \Lambda_2=-D\partial_y \left( \delta p_{(2)}^y \right) \; ,
\end{equation}
\begin{equation} 
\delta \mu_1=-D\partial_y \left( \delta p_{(2)}^x \right) \; ,
\end{equation}
\begin{equation} 
\delta \mu_2=-D\partial_x \left( \delta p_{(2)}^y \right) \; ,
\end{equation}
where $\delta p_{(2)}^A$ is given by Eq.~(\ref{deflectionangle2}), are of
second order in $h$. To lowest order, we have 
\begin{equation}
  J_{gw}=\left(   \begin{array}{cc}
1-\Lambda_{gw}          &       -\mu_{gw}       \\
-\mu_{gw}               &       1+\Lambda_{gw}   \end{array}   \right) \; .
\end{equation}
The convergence
term is absent and hence the lens action is due only to the shear. This
result was derived in Ref.~\cite{BBVarenna} using the Raychaudhuri equation 
and the optical scalars formalism, and is now recovered in the 
vector formalism.

When the determinant of $J_{gw}$ is considered, one has
\begin{equation}
\mbox{Det}(J_{gw})=1-\left( \Lambda_{gw}^2+\mu_{gw}^2 \right)+
D\left[ \partial_x \left( \delta p^x_{(2)}\right)
+\partial_y \left( \delta p^y_{(2)}\right) \right]   \; .
\end{equation}
An order of magnitude estimate as outlined in Sec.~4 gives
$\Lambda_{gw}^2$,~$\mu_{gw}^2\sim \left( D/P \right)^2 h^2>>D \,
\partial_A \left(
\delta p^A \right) \sim \left( D/P \right) h^2 $ for large $D/P$. Therefore, we
can write
\begin{equation}
\mbox{Det}(J_{gw})=1-( \Lambda_{gw}^{2}+\mu_{gw}^{2})
\end{equation}
(where the term in brackets is of second order in $ hD/P  $).

It is
to be noted that the deflection angle (\ref{deflections}) does not depend on
the frequency of the light. Gravitational waves are {\em
achromatic} lenses, like ordinary gravitational lenses. However, while the
latters do not shift the frequency of the photons propagating through them,
lensing gravitational waves do. The effect is of first order in the wave
amplitude and has been studied in detail by many authors 
\cite{Zipoy,Dautcourt,Rees,Burke2,Braginskyetal,CMB}, expecially in 
conjunction with the microwave background anisotropies discovered
by the {\em COBE} experiment.                         
In addition, gravitational waves do not rotate the polarization plane of 
the electromagnetic field, to first order \cite{FaraoniAA}. In this 
aspect, they behave like ordinary gravitational lenses. 

It is expected that the images of a distant source created by a  gravitational 
wave vary on timescales of the order of the wave period. This
could possibly be used to explain the variability of some active
galactic nuclei or active galaxies. Moreover, the details of the images 
configuration depend on the detailed form and parameters of the lensing wave, 
such as
its spatial and time profile, its duration, direction of propagation and
polarization (see \cite{Labeyrie} for an example).

\section{Order of magnitude estimates}
 
In order to apply the previous theory and the multiple images to be
detectable, the following conditions must be satified:
\begin{enumerate}
\item geometric optics holds;
\item the scale of separation between different images must not be smaller
than $ 10^{-3} $ arcseconds. In fact, structures on scales $\sim 10^{-3} $
arcseconds can be resolved with VLBI, while VLA and optical
techniques apply on larger scales\footnote{If one is not interested in the
possibility of resolving the multiple images created by the gravitational
wave, but merely in the occurrence of high amplification events, the limit
$10^{-3}$ arcseconds can be considerably relaxed.};
\item the lens must not be exceptionally rare, i.e. the rate of occurrence
of the event generating the lensing wave must not be too low;
\item in order to appreciate variability in the images induced by a lensing
gravitational wave, its period must not be too short, let us say $ P <
10^{8} $ s.
\item to ensure that the point where lensing takes place is in the
wave zone of the gravitational field, the impact parameter $ r $ must
obey the condition\setcounter{equation}{0}
\begin{equation}    \label{2}
r> \lambda_{gw}  \; .              
\end{equation}
\end{enumerate}
 
In order to satisfy 1), only electromagnetic
radiation with wavelength $ \lambda_{em} $ satisfying $ \lambda_{gw} \cdot
( \lambda_{gw}/D) <\lambda_{em}<\lambda_{gw} $ will be considered. To 
satisfy 2) note that, if $ \delta \sim h$ is the deflection angle, it must be
\begin{equation}       \label{1}
   h\geq 5\cdot 10^{-9} \;.
\end{equation}

3) depends on the particular processes generating gravitational radiation. 
Since these are almost all purely speculative, their rates of occurrence are
largely or completely unknown and we can only guess their
values. A continuous source of gravitational radiation will give rise to
a permanent lens, while a gravitational wave burst will constitute a
temporary lens.
 
The sources of gravitational waves that are most often considered in the
literature are:
\begin{itemize}
\item stellar collapse with non--spherical symmetry;
\item formation of massive black holes in active galactic nuclei;
\item neutron star collision;
\item black hole collision;
\item close binary systems;
\item black hole accretion.
\end{itemize}
The last two types of objects are {\em continuous} sources of gravitational
radiation, while the others give {\em bursts}. In addition, we will
consider the stochastic gravitational wave background, both primordial or
generated (\cite{Carr,RosiZimmerman,Matzner} and references therein).
Gravitational waves generated by a process involving a body of mass $ M $
and size $ R $ have dimensionless amplitudes (near the source) of order
\begin{equation}
    h_{S}\sim \epsilon \,\, \frac{M}{R}   \;,
\end{equation}
where the {\em efficiency} $ \epsilon $ is defined as the fraction of energy
radiated away. For processes involving neutron stars or black holes one can
assume $ M/R \sim 1/20 $ and $ M/R \sim 1 $, respectively.
 
Multiple imaging by gravitational waves is, in principle, possible, 
and it should be expected if they satisfy
the approximate condition~(\ref{roughcondition}). We examine the astrophysical
sources of gravitational radiation which are most often considered in the
literature, in conjunction with the condition (\ref{roughcondition}). 
When the event generating gravitational waves
involves neutron stars or black holes, the ordinary lensing associated to
these objects (``microlensing'' \cite{SEF})
should, in principle, be taken into account: however the separation scale
between microimages of a distant source
created by a compact object is of order $ 10^{-6} $ arcseconds, not
detectable with present techniques, while the multiple images due to the
gravitational wave may be detectable. In addition, the effects induced by
gravitational waves vary on a scale different from the typical scale of
variation of microlensing, and include a frequency shift $\delta \nu/\nu \sim
h$, which is absent in ordinary microlensing (and probably unobservable).
However, even in the situations in which lensing by gravitational waves cannot
be separated by ordinary microlensing, the former may be dominant, and it is
important to study how gravitational waves modify the microlensing phenomenon.

\begin{quote}
{\bf Stellar core collapse}
\end{quote}
The research program on collapsing homogeneous
ellipsoids by Saenz and Shapiro \cite{SS78}--\cite{SS81} has
given the expected maximal efficiences for a ``cold'' and ``hot'' equation of
state as $\epsilon \sim 10^{-2} $ and $ \epsilon \sim 10^{-4} $
respectively, for a spectrum of emitted gravitational radiation broadly
peaked between 100 Hz and 1 KHz. Taking the lower value $ \epsilon \sim
10^{-4} $ one gets $ h_{S}\sim 5 \cdot 10^{-6}$ for the wave amplitudes near
the collapsing core and the condition $ h \sim h_{S}R_{S}/r > 5\cdot
10^{-9} $ implies $ r < 10^{3}R_{S} \sim 10^{9} $ cm. On the other hand
it must be $ r > \lambda_{gw} \simeq 3\cdot 10^{7}-10^{8} $ cm; there
is a rather narrow permitted range for the impact parameter ($ r\sim 10^{8}
$ cm), that gives $ Dh/P \sim 10 $ if $ D \sim 6 \cdot 10^{15} $ cm.
 
If the late phase when the ellipsoid has settled down as a rapidly rotating
neutron star is taken into account,
it is found \cite{DetweilerLindblom} that the emitted spectrum is very
narrowly peaked ($\Delta \nu /\nu \sim 10^{-3} $) and $ \epsilon \sim
10^{-6} $ at $ \nu \sim 1$ KHz, that gives $ h_{S} \sim 5\cdot 10^{-8}
$. Condition~(\ref{1}) requires $ r< 10 R_{S}\sim 10^{7} $ cm; on the other
hand, condition~(\ref{2}) does not
allow for lensing on a relevant scale to take place in the wave zone.
 
If the core keeps bouncing, its eccentricity becomes large after a sufficient
number of 
bounces. This asymmetry \cite{SS81} makes the efficiency almost uniformly
near its maximum value for any initial period above 1 sec to several
hundred seconds \cite{Eardley}. One gets $ h_{S} \sim 10^{-2} $ at
$ \nu \sim 1 $ KHz; the condition~(\ref{1}) gives $ r< 2\cdot 10^{6}R_{S}
\sim 2 \cdot 10^{12} $ cm, while (\ref{2}) gives $ r \geq 3 \cdot 10^{7} $
cm. A rather large
range of values of the impact parameter is permitted; one obtains 
$ Dh/P \sim 10
$ if $ D \sim 10^{12} $ cm, $ r \sim 3\cdot 10^{7} $ cm, or if $ D \sim
6 \cdot 10^{16} $ cm, $ r \sim 2 \cdot 10^{12} $ cm.
 
Studies of the perturbations of pressureless spherical collapse leading to
the formation of a black hole \cite{MCP80}--\cite{CMP79} give results that
could possibly be extrapolated to larger deviations from spherical symmetry
\cite{Eardley}, getting $ \epsilon \sim 2\cdot 10^{-2} \left(
\frac{J}{M^{2}} \right) ^{4} $ at $ \nu \sim 1KHz\cdot \left(
M/10M_{\odot} \right)^{-1}$ for $ J/M^{2} \ll 1 $ (where $ J $ is
the angular momentum and $ J=M^{2} $ corresponds to a maximally rotating
Kerr black hole). Taking $ J \sim 0.1 \,M^{2} $ and $ M=10 M_{\odot} $,
it follows that $ h_{S} \sim 2\cdot 10^{-6} $ and (\ref{1}) implies
$ r< 400 R_{S}\sim 4\cdot 10^{8} $ cm,
while (\ref{2}) gives $ r\geq 3\cdot 10^{7} $ cm. A rather narrow range
of values of $ r\sim 10^{8} $ is permitted, that gives $ Dh/P \sim 10 $ if
$ D \sim 3\cdot 10^{16} $ cm.
 
\begin{quote}
{\bf Final decay of a neutron star/neutron star binary}
\end{quote}
 
Rough estimates for the final decay of a binary system composed of two
neutron stars \cite{ClarckEardley} give $ \epsilon \sim 5\cdot 10^{-3} $ at
$ \nu \leq 2-3$ KHz. Taking $ \nu \sim 500$ Hz it follows that 
$ h_{S} \sim 2.5 \cdot
10^{-4} $. (\ref{1}) gives $ r< 5\cdot 10^{4}R_{S} \sim
5\cdot 10^{10} $ cm, while $ r> 6\cdot 10^{7} $ cm due to (\ref{2}). A
large range of
values of $ r $ is permitted; we get $ Dh/P \sim 10 $ if $ D\sim 3\cdot
10^{16} $ cm, $ r \sim 5\cdot 10^{10} $ cm, or if $ D\sim 6\cdot 10^{7} $
cm, $ r \sim 6\cdot 10^{7} $ cm.
 
\begin{quote}
{\bf Black hole collisions}
\end{quote}
 
The head--on collision of two equally massive, non--rotating black holes has
been studied numerically \cite{Smarr79a}, leading to a single,
larger, black hole, with efficiency $ \epsilon \sim 7\cdot 10^{-4} $. If
the two initial black holes have nearly enough angular momentum to go into
orbit before coalescing, a formula derived from extrapolation of
perturbation theory \cite{Eardley,Smarr79a} gives $ \epsilon \sim 3\cdot
10^{-2} $. This efficiency is expected to hold for $ P \sim 1 $ s;
(\ref{1}) gives $ r < 6\cdot 10^{6}R_{S}\sim 6\cdot 10^{12} $ cm,
while (\ref{2}) implies $ r\geq 3\cdot 10^{10} $ cm. In the permitted
range of values of $ r $ the condition $ Dh/P \sim 10 $ is satisfied 
if $ D \sim 3\cdot
10^{17} $ cm, $ r\sim 3\cdot 10^{10} $ cm and if $ D\sim 6\cdot 10^{19}
$ cm, $ r\sim 10^{12} $ cm.

The case of coalescing black holes was studied also in
Ref.~\cite{Durrer}. Although multiple imaging was not considered in 
that paper, the deflection angle of light rays was computed using a 
gauge--invariant formalism.

\begin{quote}
{\bf The binary pulsar}
\end{quote}
 
The binary pulsar PSR~1913+16 (\cite{PSR1913+16} and references therein) 
is
believed to radiate gravitational waves in a continuous way, according
to the predictions of general relativity (\cite{Will} and references
therein). The estimated distance of the binary system (believed to be a
neutron star/neutron star system) is $ D \sim 5$ Kpc and
the frequency of the radiation is twice the orbital frequency (due to the
quadrupole nature of the radiation). From these values one 
obtains\footnote{The orbital period is $ \sim 2.8 \cdot 10^{4} $~s 
\cite{Will}.} $ D/P \sim 3.5\cdot 10^{7} $ and
$  Dh/P \sim 10 $ if $ h \sim 3\cdot 10^{-7} $. An estimate of the
amplitude of the waves emitted by the binary pulsar gives
\begin{equation}
h \sim \frac{\ddot{Q}}{r} \sim \frac{Ma^{2}\omega^{2}}{r}\;,
\end{equation}
where $ Q $ is the quadrupole moment, $ M $ is the mass and
$ a \sim 7\cdot 10^{10} $ cm \cite{Will} is the semimajor axis of the binary
system, so that $ h_{S}\sim
\left( a\omega /c \right)^{2}\sim 10^{-6} $. It is easy to see that
the conditions (\ref{1}) and (\ref{2}) are not compatible. Therefore
multiple imaging by the gravitational waves emitted by the binary pulsar
PSR~1913+16 is impossible.
 
\begin{quote}
{\bf The gravitational wave background}
\end{quote}
 
For the gravitational wave background, both primordial or generated 
(\cite{Carr} and references therein), one has
\begin{equation} 
h \sim \sqrt{\Omega_{gw}}\,\,\,\, \frac{P_{0}}{R}   \;,
\end{equation}
where $ P_{0} $ is the present gravitational wave period, $ R $ is the
radius of the universe and $ \Omega_{gw} $ is the cosmological density of
gravitational waves (in units of the critical density). One has 
\begin{equation}   \label{56}
\frac{Dh}{P} \sim \sqrt{\Omega_{gw}}\,\,\frac{D}{R} \;.
\end{equation}
Upper bounds on $ \Omega_{gw} $ have been set in various bands of
frequencies. Apart from
the obvious bound $ \Omega_{gw} <1 $ for every region of the spectrum, 
the limit $ \Omega_{gw}\ll 1 $ has been established in many frequency 
bands (\cite{Carr,RosiZimmerman,Matzner} and references therein). Moreover, 
$ D/R <1 $, hence it is likely that $ Dh/ P \ll 1
$ (the exact value depending on the frequency band) and one concludes 
that 
multiple imaging by the gravitational wave background is, on average,
impossible. \\

We conclude this section with an example: gravitational waves emitted in a
supernova collapse in the Virgo cluster. While the dimensionless amplitude $h_1
\sim 10^{-21}$ on the earth (at a distance $r_1 \simeq 15$~Mpc), one has $h_2
\sim h_1 r_1/r_2$ at a distance $r_2$. If $r_2 \sim 1$~pc, $h_2 \sim 1.5 \cdot
10^{-4}$ and assuming $P \simeq 10^{-3}$~s, Eq.~(\ref{lab}) tells us that
$\sqrt{f( \alpha)} Dh_2/P$ has to be greater than, or of the order unity in
order to have multiple images. This gives $ ( 1-\epsilon  )^{-1} 
\leq D_S/D_{LS} \leq \epsilon^{-1}$, where $\epsilon =4.4 \cdot 10^{-15}$, 
hence
multiple imaging is possible with a light source located virtually anywhere
beyond the Virgo cluster (of course this situation is different if one changes
the parameters $h_2$, $P$).

The amplification on a caustic is infinite in the geometric optics approximation
and is limited by geometric optics. For ordinary gravitational lenses, 
${\cal A}
\sim \left( M/ \lambda_{e.m.} \right)^{1/3}$ on a fold caustic and 
${\cal A} \sim \left( M/ \lambda_{e.m.} \right)^{1/2}$ on a cusp caustic, where
$M$ is the mass of the lens and $\lambda_{e.m.}$ is the electromagnetic
wavelength \cite{McBreenMetcalfe}. Naively, one expects a similar result 
for lensing gravitational waves, when $M$ is substituted with $( h/P)^2
l^3$, where $l$ is the size of the gravitational wave packet. However, a
realistic computation on the lines of \cite{SEF} is much more complicated and
requires the use of precise models that we have not considered here. 

\section{Discussion and conclusions}

Gravitational waves affect the propagation of light; this is expected for any
gravitational field. The interaction between light and gravitational waves in
general relativity has
been considered by many authors \cite{Zipoy}--\cite{Braginskyetal}. 
Recently, the attention was focussed upon the
contribution of gravitational waves to the microwave background anisotropies
discovered by the {\em COBE} experiment \cite{CMB}, and on the possibility 
of detecting 
astrophysically generated (as opposed to primordial) gravitational waves
through their effects on light rays \cite{Labeyrie}--\cite{Durrer}. The 
possibility of gravitational waves
superimposed to an ordinary gravitational lens was also considered 
\cite{McBreenMetcalfe}--\cite{FriemanHarariSurpi}. 
The generalization of the Fermat principle to
non--stationary spacetimes \cite{Fermat} allows one to approach 
the problem using a modified version of the vector formalism employed 
for ordinary 
gravitational lenses \cite{FaraoniApJ}. An approximate condition
(Eq.~(\ref{roughcondition})) analogous to that 
holding for ordinary gravitational lenses must be satisfied in order to 
create multiple 
images of a distant light source. The astrophysical sources of gravitational
waves most often considered in the literature have been examined. Certain
astrophysical sources are shown to
satisfy the approximate condition for the creation of multiple images, due 
to the balance between the large values of the
distances involved and the small values of the gravitational wave amplitudes.
Another relevant fact is that lensing can take place in regions relatively 
close to the sources of
gravitational waves, where the wave amplitudes are larger than those expected
in the Solar System. The gravitational wave background, instead, is 
unlikely to produce multiple images. As a conclusion, 
multiple imaging of a
distant source by gravitational waves is possible 
in the favourable situations examined, including the collapse of
stellar cores, the final decay of neutron star/neutron star binaries and
black hole collisions. Unfortunately, these events (in particular black hole 
collisions) are very rare and the probability of
observing the phenomenon is extremely low, mainly because the duration of the
multiple images is limited to the period of intense emission of
gravitational radiation. Continuous sources of gravitational waves like the
binary pulsar PSR 1913+16 would give a much higher probability, but
unfortunately they emit gravitational waves too weakly.

It is to be noted that the limit $\delta \geq 10^{-3}$ arcseconds in 
condition~2) of Sec.~6 can be considerably relaxed if we do not require that
the multiple images created by gravitational waves be resolved. The high
amplification events associated to the presence of multiple images would
still constitute an interesting and observable phenomenon even if the multiple
images are not resolved. In fact, this situation occurs in ordinary
microlensing by stars or planets: the typical scale of separation of
microimages is $10^{-6}$ arcseconds or less, yet the phenomenon has been
observed \cite{Aubourgetal,Udalskietal}. Moreover, it was not necessary to
consider cosmological distances $D$ in order to have 
an appreciable probability
for microlensing: the phenomenon was observed in the Large Magellanic Cloud
\cite{Aubourgetal} and in our galaxy \cite{Udalskietal}.

Further work is necessary to assess precisely the probability of
observing multiple images created by gravitational waves from 
astrophysical sources, 
and the details of the phenomenon. Contrarily to the conclusions of
Ref.~\cite{Labeyrie}, we are not very optimistic on the
probability of observing such an event, but our results show that this 
possibility deserves some attention. In addition, we restricted our attention
to the creation of multiple images, while Ref.~\cite{Labeyrie} 
include other aspects of lensing. It is also useful to remember that, before 
the discovery of
the first gravitational lens system in 1979, gravitational lenses were 
considered mere speculations not occurring in the real world. Today, not 
only the existence of gravitational lenses is universally
acknowledged, but gravitational lensing has become one of the most active 
and promising fields of astronomy.

\section*{Acknowledgments}

The author is deeply indebted to B. Bertotti for suggesting the
possibility of lensing by gravitational waves and the 
approach used in this paper, and for helpful discussions. 
Thanks are due also to G.F.R. Ellis for stimulating discussions and to 
a referee for suggestions leading to improvements in the manuscript.

\section*{Appendix}

Here we provide an independent proof of the equation 
$\partial_A ( \delta p^A )=0$, which was found also in
Ref.~\cite{BBVarenna}. From Eq.~(\ref{usedintheapp}) and from the normalization
$p_{\mu}p^{\mu}=0$ we obtain, to first 
order,\def\theequation{A.\arabic{equation}}\setcounter{equation}{0}
\begin{equation}  \label{A1}
2\eta_{\mu\nu} p_{(0)}^{\mu} \delta p^{\nu}+h_{\mu\nu} p_{(0)}^{\mu}
p_{(0)}^{\nu}=0 \; ,
\end{equation}
and
\begin{equation}            \label{A2}
\delta p^0 - \delta p^3 =\frac{1}{2} \left( h_{00} +2h_{03}+h_{33} \right) +
{\mbox O}(h^2) \; .
\end{equation}
From Eq.~(\ref{deflectionangle}) it follows 
that $\partial_3 ( \delta p^{\mu})=0$ and Eq.~(\ref{A2}) gives 
\begin{equation}    \label{A3}
\partial_3 \left( h_{00}+2h_{03}+h_{33} \right)=0 +{\mbox O}(h^2) \; .
\end{equation}
Equation~(\ref{deflectionangle}) gives $\delta p^3=0$ and from Eq.~(\ref{A2})
it follows that
\begin{equation}            \label{A4}
\delta p^0 =\frac{1}{2} \left( h_{00} +2h_{03}+h_{33} \right) +{\mbox O}(h^2)
\; .  
\end{equation}
By setting $f\equiv ( h_{00}+2 h_{03}+h_{33} ) /2 $, it follows from 
Eq.~(\ref{deflectionangle}) that $\partial_3 f=-\partial_0 f$ and
\begin{equation}   \label{1star}
\partial^2_{00}f-\partial^2_{33}f =0 
\end{equation} 
along the photon trajectory. Using the linearized Einstein equations
for gravitational waves propagating outside their sources, 
\begin{equation}
\Box \bar{h}_{\mu\nu}=0 \; ,
\end{equation}
\begin{equation}
\partial^{\nu} \bar{h}_{\mu\nu}=0 \; ,
\end{equation}
where $ \bar{h}_{\mu\nu} \equiv  h_{\mu\nu}-\eta_{\mu\nu} h/2$, one obtains
\begin{equation}  \label{AA}
\Box h_{00}=-\Box h_{33}=-\frac{1}{2} \Box h \;\;\;\;\;, \;\;\;\;\; 
\Box h_{03}=0 \; .
\end{equation}
From Eqs.~(\ref{AA}) one obtains 
\begin{equation}   \label{Boxf}
\Box f=0
\end{equation} 
and
\begin{equation}
\partial_{\mu} {\left( \delta p^{\mu} \right)=\frac{1}{2} \int_S^O dz \left( 
h_{00}+2h_{03}+h_{33} \right)^{,\mu}}_{,\mu}=0 +{\mbox O}(h^2) \; .
\end{equation}
The two--dimensional divergence $\partial_A \left( \delta p^A \right)$ is given
by
\begin{equation}   
\partial_A \left( \delta p^A \right)=\int_S^O dz \, {f^{,A}}_{,A}=
\int_S^O dz \left( \Box f+\frac{\partial^2 f}{\partial t^2}-   
\frac{\partial^2 f}{\partial z^2}\right) =0 \; , \label{AAA}
\end{equation}
where the last equality follows from Eqs.~(\ref{1star}) and (\ref{Boxf}).

\end{document}